\documentstyle[12pt]{article}
\input epsf
\newcommand {\al}   {\alpha}       \newcommand {\bt}  {\beta}
\newcommand {\g }   {\gamma}       \newcommand {\G }  {\Gamma}
\newcommand {\dl}   {\delta}       
\newcommand {\z }   {\zeta}

\newcommand {\f }   {\varphi}      \newcommand {\h }  {\chi}
         \newcommand {\om}  {\omega}
\newcommand {\pl}   {\partial}     
\begin{document}
\title  {Scattering on dislocations and cosmic strings\\
         in the geometric theory of defects}
\author {M.~O.~Katanaev\thanks{E-mail: katanaev@mi.ras.ru}
         and I.~V.~Volovich\thanks{E-mail: volovich@mi.ras.ru}\\ \\
         \it Steklov Mathematical Institute,\\
         \it Gubkin St., 8, Moscow, 117966, Russia}
\date   {12 May 1998}
\maketitle
\begin{abstract}
          We consider scattering of elastic waves on
          parallel wedge dislocations in the geometric theory of
          defects or, equivalently, scattering of point particles
          and light rays on cosmic strings.
          Dislocations are described as torsion
          singularities located on parallel lines, and trajectories
          of phonons are assumed to be the corresponding extremals.
          Extremals are found for arbitrary distribution of the
          dislocations in the monopole, dipole, and quadrupole
          approximation and the scattering angle is obtained.
          Examples of continuous distribution of wedge and edge
          dislocations are considered. We have found that for
          deficit angles close to $-2\pi$ a star located behind
          a cosmic string may have any even number of images,
          $2,4,6,\dots$
          The close relationship between dislocations
          and conformal maps is elucidated in detail.
\end{abstract}

\newpage
\section{Introduction}
Geometric description of defects in solids has attracted continuous
interest in the last years. The central role here is played by torsion.
This geometric notion was introduced by E.~Cartan \cite{Cartan22} (see
also \cite{Cartan86}) and, quite interesting, by analogy with elastic
media. Later the relation of torsion to dislocations was noticed
\cite{Kondo52,BiBuSm55} (see reviews [5-7]),
\nocite{Kroner81,KadEde83,Kleine89}
and disclinations in spin glasses or liquid crystals were found to
correspond to nontrivial curvature of a $SO(3)$-connection \cite{DzyVol78}.
Elastic media with a spin structure (Cosserat media \cite{CosCos09})
containing arbitrary distribution of dislocations and disclinations
can be consistently described in the framework of general Riemann--Cartan
geometry where torsion and curvature are identified with surface densities
of Burgers and Frank vectors, respectively \cite{KatVol92} (a similar
approach was discussed in \cite{Madore96}). Generalizations of gravity
models to Riemann--Cartan and general affine geometry are reviewed
in \cite{HeMcMiNe95}. It should be stressed that geometric description
treats single as well as continuously distributed defects in a unified
manner.

The advantage of geometric description of defects in solids is twofold.
First, in contrast to the ordinary elasticity theory this approach
provides an adequate language for continuous distribution of defects.
Second, a mighty mathematical machinery of differential
geometry clarifies and simplifies calculations.

From a mathematical point of view geometric theory of defects in solids
and the theory of gravity with torsion in the Euclidean formulation
are the same models. The relation between $N$ point particle solution
of three dimensional gravity [13--15]
and wedge dislocations was first analyzed in \cite{Holz88,Holz92}
\footnote{There wedge dislocations are called disclinations.
The last name is also used in describing defects in a spin
structure of a media. The geometric theory describes both
types of defects, and therefore we reserve the name disclination for a
defect in a spin structure.}. (See also \cite{Kohler95A,Kohler95B}.)
In \cite{KatVol92} we proposed a three dimensional action for describing
defects in solids and showed that description of an arbitrary distribution
of static parallel dislocations naturally arises from an action principle.
This solution defines nontrivial geometry on $R^3$ manifold of zero
curvature with $N$ removed parallel lines corresponding to torsion
singularities. Each of the removed lines defines a conical singularity
in the perpendicular plane. In spite of the fact that the curvature is
zero the whole space is not a manifold with absolute parallelism
because parallel transport of a vector around conical defect produces
rotation on a deficit angle. A gauge approach to the theory of defects
in solids and cosmic strings is adopted in \cite{PunSol97}.
Recently a series of articles dealing with the physical properties of
the $N$ point particle solution appeared [21--26].

In the present paper we analyze scattering of elastic waves on parallel
wedge dislocations. In the eikonal approximation or geometric optics
\cite{LanLif62} elastic deformations described by the wave equation
move along extremals. By analogy with motion of photons in
electrodynamics it is natural to assume that extremals are
trajectories of phonons in elastic media with defects
\footnote{An alternative point of view is adopted in
\cite{Kleine97} where a point particle is assumed to move along geodesic
(there extremals and geodesics are called geodesics and autoparallels,
respectively).}. In this way analysis of extremals yields a complete
picture of the scattering
of phonons on dislocations. In the case under consideration the
scattering problem is not the standard one because in general the space
is not asymptotically flat. For example, the notion of a falling beam of
parallel moving particles depends on the distance to the dislocation,
and the standard definition of a cross section does not work. At the
same time one can easily define the scattering angle considering
asymptotics of a particle trajectory. For one wedge dislocation
we derive the following scattering angle
\begin{equation}                                        \label{escata}
  \h=\frac{\pi\theta}{1+\theta},
\end{equation}
where $2\pi\theta$ is the deficit angle that is the angle of the wedge
which is removed $(-1<\theta<0)$ or added $(0<\theta)$ to the media.

Extremals for one edge dislocation were analyzed perturbatively in
\cite{Moraes96}. Recently extremals were analysed for one dispiration
defect containing a wedge dislocation as a special case \cite{dePaMo98}.
In the present paper we extend these results and analyze
extremals nonperturbatively for arbitrary distribution of wedge parallel
dislocations. In fact, a general explicit expression for an extremal is
obtained. Already for one wedge dislocation we obtain a very rich and
interesting scattering picture of phonons. In particular, if the
deficit angle is close to $-2\pi$ or $\theta\approx-1$ then a phonon
makes several turns around the dislocation before going to infinity.
If dislocations are more concentrated in a compact domain of the
perpendicular plain than at large distances one may use perturbation
theory. In this way we analyzed monopole, dipole, and quadrupole
contributions. The dipole approximation coincides with the
scattering on an edge dislocation.

The spatial part of the metric for a static cosmic string
\cite{VilShe94,HinKib95} coincides with the metric for a wedge
dislocation although it was obtained in a quite different manner
by solving four dimensional Einstein equations for a cosmic string
in the linear approximation. Therefore scattering of elastic waves
on dislocations is the same as scattering of point particles or
photons on cosmic strings. We derive a general formula for the
angular separation between images of a star located behind a cosmic
string. For small angles the expression coincides with the result
obtained earlier, and there are only two images. If the deficit
angle approaches $-2\pi$ then the number of images increases
because an observer sees the light rays which make no rotation
around the string, one rotation, two rotations, and so on.

The results are applicable to a radial disgyration defect in
$^3$He-A, described by a conical singularity, too \cite{Volovi97}.

In Section \ref{sgenfr}, a general framework for the analysis of extremals
is briefly described. It is applied to extremals in the case of one
wedge dislocation in Section \ref{swedge}. In Section \ref{sangse}
we calculate the angular separation between the images of a star
located behind a cosmic string. The analysis of extremals is extended
to arbitrary distribution of wedge dislocations in Section \ref{sexwen}.
In Section \ref{sconma} we show that parallel linear dislocations
are described by the conformal map of the Christoffel--Schwarz type.
The dipole approximation corresponding to pure edge dislocation is
considered in detail in Section \ref{sdipap}. In Section \ref{squdap}
we study the quadrupole approximation. Two simple examples of continuous
distribution of dislocations are considered in Section \ref{scondi}.
\section{A general framework                           \label{sgenfr}}
We suppose that elastic media with dislocations is an $R^3$ manifold
with some punctures and/or removed lines corresponding to the cores of
dislocations and a given Riemann--Cartan geometry \cite{KatVol92}.
An arbitrary curvilinear coordinate system is denoted by $x^\mu$,
$\mu=1,2,3$. A Riemann--Cartan geometry is defined by metric and torsion
or, equivalently, by triad and $SO(3)$-connection. Here we use the more
convenient second set of variables called Cartan variables.
Pure dislocations correspond to zero curvature
\begin{equation}                                     \label{ecurth}
  R_{\mu\nu}{}^{ij}=\pl_\mu\om_\nu{}^{ij}
  -\om_\mu{}^{ik}\om_{\nu k}{}^j-(\mu\leftrightarrow\nu)=0,
\end{equation}
but nontrivial torsion. Here $\om_\mu{}^{ij}=-\om_\mu{}^{ji},$
$i,j,\dots=1,2,3,$ is an $SO(3)$-connection considered as an
independent variable. For zero curvature it is a pure gauge and
can be always set to zero at least locally by means of local $SO(3)$
rotation. This can be done in domains with the trivial fundamental group
which are spaces with absolute parallelism. If the space contains
singularities and one has to remove infinite or closed lines from
$R^3$ then this property is violated but may be restored by making
appropriate cuts. A triad field $e_\mu{}^i$
defines a metric $g_{\mu\nu}$ in the usual way
\begin{equation}                                     \label{emettr}
  g_{\mu\nu}=e_\mu{}^i e_\nu{}^j\dl_{ij},~~~~~~
  \dl_{ij}=diag(+++).
\end{equation}
In our case torsion tensor,
$$
  T_{\mu\nu}{}^i=\pl_\mu e_\nu{}^i-\om_\mu{}^{ij}e_{\nu j}
  -(\mu\leftrightarrow\nu),
$$
is entirely defined by the triad because the $SO(3)$-connection is set
to zero.

Of course, for a given metric (\ref{emettr}) one can construct
Christoffel's symbols and the corresponding curvature as if torsion equals
zero. It will be not the same curvature as given by (\ref{ecurth}) and
is nontrivial in the presence of dislocations.

We assume that phonons in the media with dislocations move along extremals
defined by the metric (\ref{emettr}). Extremals $x^\mu(t)$ satisfy the
well known set of equations
\begin{equation}                                      \label{eextre}
  \ddot{x}^\mu=-\G_{\nu\rho}{}^\mu\dot{x}^\nu\dot{x}^\rho,
\end{equation}
where $\G_{\nu\rho}{}^\mu$ denotes Christoffel's symbols
(the Levi--Civita connection) constructed from the metric (\ref{emettr}).
Note that extremals coincide with geodesics in Riemann--Cartan geometry
if and only if torsion is zero or antisymmetric in all three indices.
In three dimensions the canonical parameter $t$ is interpreted as the time.
Indeed, extending three-dimensional space to four-dimensional space-time,
$x^\mu\rightarrow(x^0,x^\mu)$, and assuming the following form of the
four-dimensional metric
$$
  \left(\begin{array}{cc}-1&0\\0&g_{\mu\nu}\end{array}\right),
$$
one finds the fourth equation for an extremal
$$
  \ddot{x}^0=0~~~~~~\Leftrightarrow~~~~~~x^0=at+b,~~a,b=const.
$$
Thus up to a linear transformation the canonical parameter coincides
with the real time $x^0$.

Analysis of extremals becomes simpler if one notices that there is always
one integral to equations (\ref{eextre})
\begin{equation}                                        \label{exleta}
  g_{\mu\nu}\dot{x}^\mu\dot{x}^\nu=C_0=const>0,
\end{equation}
which means that the length of the tangent vector or the square of
the velocity of a phonon (the kinetic energy) remains constant.
Another simplification occurs if the metric admits a Killing vector
field $k^\mu$,
$$
  \nabla_\mu k_\nu+\nabla_\nu k_\mu=0,
$$
where $\nabla_\mu$ denotes the covariant derivative with respect to
Christoffel's symbols. In this case there exists another integral
\begin{equation}                                        \label{exkive}
  g_{\mu\nu}k^\mu\dot{x}^\nu=C_1=const,
\end{equation}
which is related to the symmetry of the space.
In two dimensions the existence of one Killing vector field is enough
for the complete analysis of extremals because one has two integrals
(\ref{exleta}), (\ref{exkive}) for two unknown functions. In higher
dimensions complete analysis requires the existence of more Killing
directions or some other techniques. The latter is also required in two
dimensions in the absence of a Killing vector field.

Another approach to the analysis of extremals is provided by the
action principle. It is well known that equations (\ref{eextre})
follow from the Lagrangian
\begin{equation}                                     \label{elagex}
  L=\frac12 g_{\mu\nu}\dot{x}^\mu\dot{x}^\nu,
\end{equation}
which equals the kinetic energy of a point particle of unit mass
moving in a space with nontrivial geometry. Of course, here the
canonical parameter and real time are naturally identified. We see
that both phonons and point particles move along the same trajectories.
The difference is only in the velocity of movement. If a particle
carries charge or spin which directly interacts with the media then
one has to add a potential term to the Lagrangian (\ref{elagex}).
In that case a trajectory will differ from an extremal. Although the
potential term in (\ref{elagex}) is absent the motion of a particle
is nontrivial because the metric explicitly depends on a position.
Then equations (\ref{eextre}) are nothing more than the Newton law
with the force quadratic in the velocity. Due to this circumstance
the motion of phonons in the presence of a dislocation differs
drastically from the Newtonian motion of a point particle in a
potential field. For example, we shall find later a new type of
closed trajectories.

If there is not enough symmetry for the complete analysis of extremals
then the Hamilton--Jacobi equation may be useful. The Lagrangian
(\ref{elagex}) gives rise to the Hamiltonian
$$
  H=\frac12 g^{\mu\nu}p_\mu p_\nu,
$$
where $g^{\mu\nu}$ is the inverse metric and
$p_\mu$$=\pl L/\pl\dot{x}^\mu$ is the canonical momenta. Then the
Hamilton--Jacobi equation for the characteristic function $S(x^\mu,t)$
takes the form
\begin{equation}                                        \label{ehamja}
  \frac{\pl S}{\pl t}+\frac12 g^{\mu\nu}
  \frac{\pl S}{\pl x^\mu}\frac{\pl S}{\pl x^\nu}=0,
\end{equation}
or
$$
  S(x^\mu,t)=W(x^\mu)-Et,~~~~~~E=const,
$$
where
\begin{equation}                                         \label{ehajas}
  \frac12 g^{\mu\nu}\frac{\pl W}{\pl x^\mu}\frac{\pl W}{\pl x^\nu}=E.
\end{equation}
Here $E$ denotes the total (in our case kinetic) energy of a phonon.
This equation is solved in Section~\ref{sexwen} for arbitrary distribution
of $N$ wedge dislocations characterized by one Killing direction.
\section{Extremals for a wedge dislocation             \label{swedge}}
\begin{figure}[htb]
 \begin{center}
 \leavevmode
 \epsfxsize=6cm
 \epsfbox{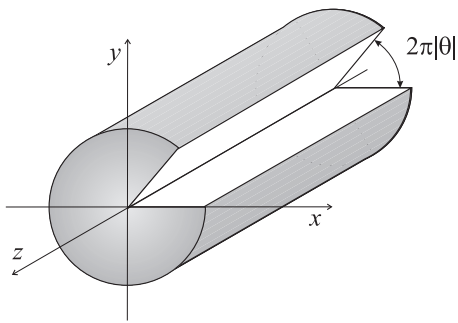}
 \end{center}
 \caption{A wedge dislocation with the deficit angle $2\pi\theta$.
          For negative $\theta$, $-1<\theta<0$, the wedge is removed
          from the media. \label{fwedge}}
\end{figure}

Let us analyze extremals in the simplest case of the media with
one wedge dislocation, characterized by the deficit angle $\theta$
normalized on $2\pi$, so that the real angle is equal to $2\pi\theta$
(see Fig.\ref{fwedge}). In this section we derive the formula
(\ref{escata}) for the scattering angle.
The $x^3$ axis is chosen along the core of dislocation.
The corresponding metric has the form \cite{KatVol92}
\begin{equation}                                        \label{emetth}
  ds^2=dl^2+(dx^3)^2,
\end{equation}
where the nontrivial two-dimensional part of the metric in polar
coordinates reads 
\begin{equation}                                        \label{emetwe}
  dl^2=r^{2\theta}(dr^2+r^2d\f^2).
\end{equation}
This is the well known form of the metric for a conical singularity with
the deficit angle $2\pi\theta$. It is not asymptotically flat, and 
particles at infinity cannot be considered as free. They feel a wedge 
dislocation even at infinite distance. Christoffel's symbols 
corresponding to the metric (\ref{emetwe}) are easily calculated,
\begin{eqnarray*}
  \G_{rr}{}^r&=&\frac\theta r,
\\
  \G_{\f\f}{}^r&=&-(1+\theta)r,
\\
  \G_{r\f}{}^\f=\G_{\f r}{}^\f&=&\frac{1+\theta}r,
\end{eqnarray*}
the others being identically zero.
Then equations for an extremal (\ref{eextre}) reduce to
\begin{eqnarray}                                        \label{extrer}
  \ddot{r}&=&-\frac\theta r\dot{r}^2+(1+\theta)r\dot{\f}^2,
\\                                                      \label{extref}
  \ddot{\f}&=&-2\frac{1+\theta}r\dot{r}\dot{\f},
\\                                                      \label{extrez}
  \ddot{x^3}&=&0.
\end{eqnarray}
The last equation is the consequence of the translational symmetry along
the $x^3$ axis, and shows that phonons move along it with a constant
velocity. Therefore the scattering reduces to the two-dimensional problem
in the $r,\f$ plane. The metric for a wedge dislocation possesses a
rotational symmetry around the $x^3$ axis. The corresponding Killing
vector field in polar coordinates has a simple form
$$
  k^\mu=\left(\begin{array}{c}0\\1 \end{array}\right).
$$
Thus integrals (\ref{exleta}) and (\ref{exkive}) for a wedge dislocation
reduce to
\begin{eqnarray}                                        \label{extfin}
  r^{2\theta}\dot{r}^2+r^{2(1+\theta)}\dot{\f}^2&=&C_0>0,
\\                                                      \label{extsin}
  r^{2(1+\theta)}\dot{\f}=C_1.
\end{eqnarray}
They have direct physical interpretation to be discussed later.
For nonradial motion one easily finds the general form of an extremal
\begin{equation}                                        \label{eforwe}
  \left(\frac r{r_m}\right)^{2(1+\theta)}\sin^2[(1+\theta)(\f+\f_0)]=1,
\end{equation}
where
\begin{equation}                                        \label{econrm}
  r_m=\left(\frac{C_1^2}{C_0}\right)^{\frac1{2(1+\theta)}},~~~~~~
  \f_0=const.
\end{equation}
We see that the form of an extremal is parametrized by two arbitrary
constants. The constant $r_m$ is positive and defines the length scale,
while the constant $\f_0$ corresponds to rotational symmetry around
the core of dislocation and may be arbitrary. Formula (\ref{eforwe})
describes an infinite number of disconnected branches depending on the
range of the angle. One branch corresponds to the variation of the
$\sin$ argument from $n\pi$ to $(n+1)\pi$, $|n|=0,1,\dots$ The radius
is bounded from below $r_m<r<\infty$, and every branch starts and ends
at infinity. For $n=0$ the angle $\f+\f_0$ varies from $0$ to
$\pi/(1+\theta)$. For positive $\theta$ the variation of the angle
is less than $\pi$, and a phonon is repelled from the dislocation.
For negative $\theta$ phonons are attracted, and can make several
turns around the dislocation. Suppose the extremal makes exactly $m$
rotations. Then the angle $\f+\f_0$ varies from $0$ to $2\pi m$, and
the corresponding deficit angle $\theta_m$ is defined by the equation
$$
  (\theta_m+1)2\pi m=\pi.
$$

\begin{figure}[htb]
 \begin{center}
 \leavevmode
 \epsfxsize=12cm
 \epsfbox{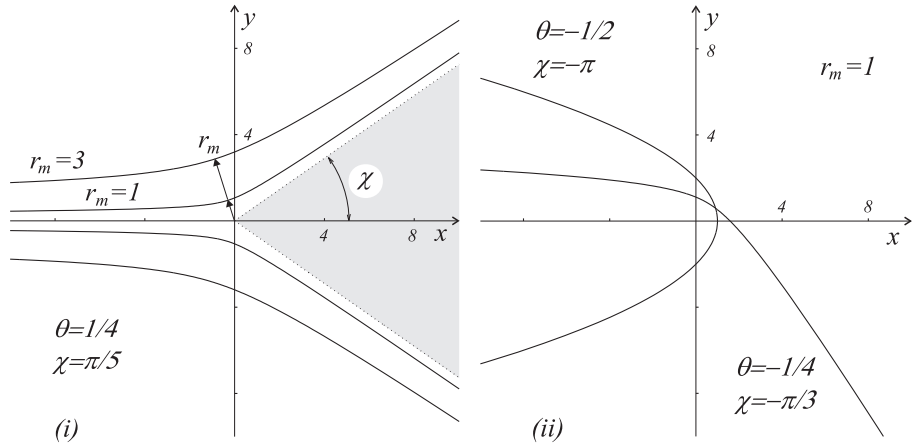}
 \end{center}
 \caption{Extremals for a wedge dislocation with $\theta>0$ {\it(i)}
          and $-1/2\le\theta<0$ {\it(ii)}. In {\it(i)} we draw two
          extremals and their reflections for the same $\theta$ but
          different $r_m$. In {\it (ii)} we draw two extremals for the
          same $r_m$ but different $\theta$. For $\theta=-1/2$ an
          extremal makes one turn around the dislocation.
          \label{fexwpo}}
\end{figure}

\begin{figure}[htb]
 \begin{center}
 \leavevmode
 \epsfxsize=12cm
 \epsfbox{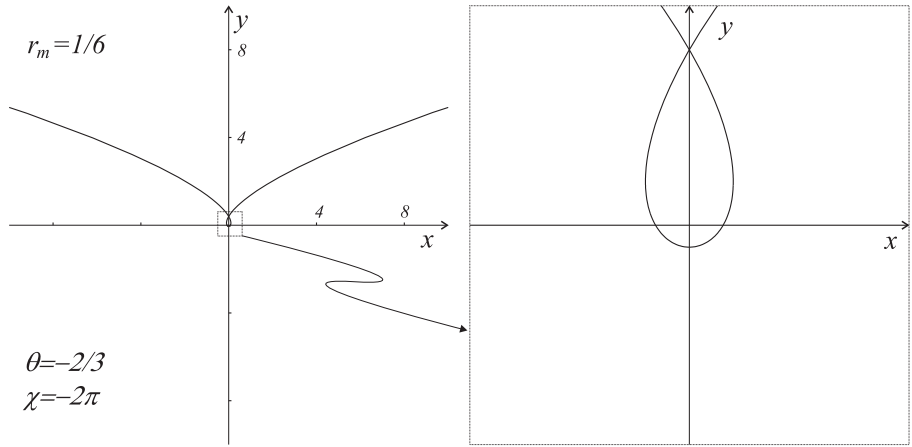}
 \end{center}
 \caption{An extremal for $\theta=-2/3$
          makes one turn around the dislocation and then goes forward
          in the same direction.
                                                       \label{fwede2}}
\end{figure}

\begin{figure}[htb]
 \begin{center}
 \leavevmode
 \epsfxsize=12cm
 \epsfbox{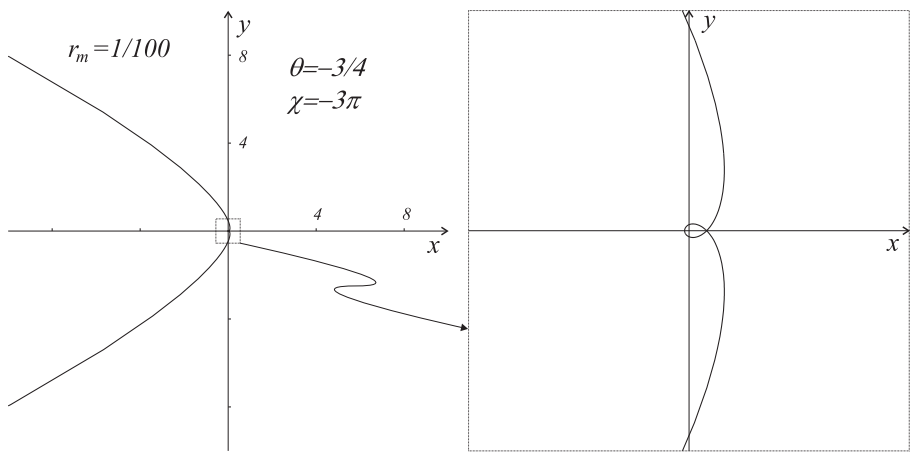}
 \end{center}
 \caption{An extremal for $\theta=-3/4$ makes two turns around the
          wedge dislocation.
                                                       \label{fwede3}}
\end{figure}

\begin{figure}[htb]
 \begin{center}
 \leavevmode
 \epsfxsize=12cm
 \epsfbox{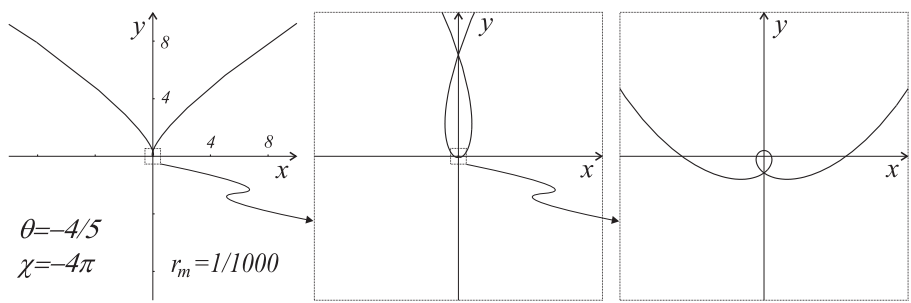}
 \end{center}
 \caption{An extremal for $\theta=-4/5$ makes two turns and a half
          around the wedge dislocation before going to infinity.
                                                       \label{fwede4}}
\end{figure}

Extremals for a wedge dislocation are drawn in
Figs.\ref{fexwpo}--\ref{fwede4}
for different values of $\theta$ and $\f_0=\pi$
\footnote{All extremals in this and the following sections are drawn
numerically, and the scale is shown in the drawings.}.
In the case $\theta>0$ the extra wedge of media of angle $2\pi\theta$
is added to the media.
For positive $\theta$
every extremal has two asymptotes going through the center and intersecting
at the scattering angle $\h$ (see Fig.\ref{fexwpo}{\it (i)}).
The scattering angle is equal to
$\pi-\triangle\f$ where $\triangle\f$ is the angle between two asymptotes
for an extremal defined by equation (\ref{eforwe}),
$(1+\theta)\triangle\f=\pi$. In this way one immediately gets the expression
$$
  \h=\pi-\frac\pi{1+\theta}=\frac{\pi\theta}{1+\theta}
$$
which coincides with (\ref{escata}). The scattering angle has clear
geometric interpretation. It equals one half of the deficit
angle times the compression coefficient
$$
  \h=\frac{2\pi\theta}2\times\frac1{1+\theta}.
$$
It means that the wedge dislocation is made by addition of uncompressed
media and afterwards the media compresses. If one added the compressed
wedge of the media then the compression coefficient would be absent.
If $\theta$ goes to infinity then any extremal returns back from the
dislocation along the same line. If the dislocation is absent, $\theta=0$,
then there is no scattering and $\h=0$. It is quite interesting that
for positive $\theta$ there are points which cannot be joined by an
extremal. For example, a point on the negative half of the $x$-axis
cannot be joined by an extremal with arbitrary point lying in the wedge
of inserted media (the shaded region in Fig.\ref{fexwpo}{\it (i)}).
The reason for this is the conical singularity at the origin of
coordinate system.

For negative $\theta$ the wedge is removed from the media. In this
case extremals have no asymptotes, but the scattering angle is still
defined by the equation (\ref{escata}). In Fig.~\ref{fexwpo} {\it(ii)}
we show two extremals for different $\theta$ but the same $r_m=1$ which
have up to one turn around the dislocation, one turn corresponding to
$\theta=-1/2$ when half of the media is removed. If $-1<\theta<-1/2$
then the phonon makes several turns around the dislocation before going
to infinity as shown in Figs.~\ref{fwede2}-\ref{fwede4}.
This kind of trajectories is not specific only to the dislocations.
Similar extremals are met already in the Schwarzschild space-time
\cite{Chandr83}. These trajectories produce multiple images of a
star located behind the singularity.

At infinity $r\rightarrow\infty$ equations (\ref{extfin}), (\ref{extsin})
yield
\begin{equation}                                        \label{extinf}
  \frac1{1+\theta}r^{1+\theta}=\sqrt{C_0}t+const.
\end{equation}
It means that for $\theta>-1,$ a particle moves to space infinity in
infinitely long time, and the manifold is complete at $r\rightarrow\infty$
(any extremal can be continued to infinite value of the canonical
parameter in both directions). Later we show that the radial extremals
can be continued through the singularity and become complete too. In
this sense the whole manifold is complete. Unphysical region $\theta\le-1$
corresponding to the case when one removes more than the whole media leads
to an incomplete manifold which should be extended.

Physical interpretation of the first integrals (\ref{extfin}),
(\ref{extsin}) is clear from the following consideration. In the primed
coordinate system,
$$
  r'=\frac1{1+\theta}r^{1+\theta},~~~~~~\f'=(1+\theta)\f,
$$
the line element (\ref{emetwe}) becomes flat
$$
  dl^2=dr'^2+r'^2d\f'^2.
$$
This means that in these coordinates phonons move along straight
lines, and the energy, momentum, and angular momentum are conserved.
The kinetic energy of a phonon of unit mass is easily calculated
$$
  E'=\frac12(\dot{r}^{\prime2}+r'^2\dot{\f}^{\prime2})=\frac12C_0.
$$
So the first integral (\ref{extfin}) is proportional to the kinetic
energy or the square of the velocity. The momentum is defined by $C_0$
and the direction of the trajectory. The calculation of the angular
momentum,
$$
  L'=r'^2\dot{\f}'=\frac1{1+\theta}C_1,
$$
yields physical interpretation for the other integration constant.
It is defined by the initial condition of the system and is arbitrary.
The constant of motion $C_0$ exists for the wave equation of acoustic
waves \cite{LanLif62} and is different for longitudinal and transverse
phonons. This is the constant distinguishing trajectories of different
waves. One may assume that point particles also move along extremals
if they do not carry charge or spin which directly interact with the
media. Then the constant $C_0$ is parametrizing the type of a phonon
and a point particle. For a point particle it is defined by the initial
condition and may be arbitrary. The form of an extremal (\ref{eforwe})
is defined by the constant (\ref{econrm}) depending both on $C_0$
and $C_1$. The constant $C_1$ is arbitrary and therefore phonons
and point particles move along the same trajectories, the only
difference being the velocity of movement.

Suppose that the external observer examines the scattering using
some device which does not interact with the media. Then he
observes the scattering on dislocation in polar coordinates
$r,\f$ which are flat for his device. For him the existence of
dislocation exhibits itself as the bending of phonon trajectories.

Let us compute physical characteristics of the scattering phonon as
measured by an external observer for whom a dislocation appears through
the properties of a phonon. The observed kinetic energy,
\begin{equation}                                        \label{eobkie}
  E=\frac12(\dot{r}^2+r^2\dot{\f}^2)=\frac12C_0r^{-2\theta},
\end{equation}
is positive and explicitly depends on the distance to the dislocation. 
It decreases and increases for positive and negative deficit angle $\theta$, 
respectively, as the phonon approaches the dislocation. At the infinity the 
energy tends to zero for positive $\theta$ and to infinity for negative 
$\theta$:
\begin{eqnarray*}
  &&\theta>0:~~~~E\rightarrow0,~~~~r\rightarrow\infty,
\\
  &&\theta<0:~~~~E\rightarrow\infty,~~r\rightarrow\infty.
\end{eqnarray*}
Of course in a real world the energy cannot increase to infinity and
is bounded by the finite size of the bodies.
The radial and angular components on the momentum have the form
\begin{eqnarray}                                        \label{eradmo}
  p_r&=&\dot{r}
  =\sqrt{C_0}r^{-2\theta-1}\sqrt{r^{2(1+\theta)}-r_m^{2(1+\theta)}},
\\                                                      \label{eangmo}
  p_\f&=&r\dot{\f}=C_1r^{-2\theta-1}.
\end{eqnarray}
The absolute value of the angular component increases and decreases
as the phonon goes closer to the dislocation for $\theta>-1/2$ and
$-1<\theta<-1/2$, respectively. For $\theta=-1/2$ it does not depend
on the distance. The behavior of the absolute value of the radial
component is more complicated. For $-1<\theta<0$ it grows up monotonically
from zero to infinity as the radius grows from $r_m$ to $\infty$.
For $\theta>0$ it is zero both at $r_m$ and infinity and has one maximum
at
$$
  r=r_m\left(\frac{1+2\theta}{2\theta}\right)^{\frac1{2(1+\theta)}}.
$$
The angular momentum as measured by an external observer,
$$
  L=r^2\dot{\f}=C_1r^{-2\theta},
$$
also depends on the distance.

Radial extremals must be treated separately. Equation (\ref{extsin})
for $\f=const$ and $C_1=0$ is satisfied, and equation (\ref{extsin})
can be easily integrated
$$
  r^{1+\theta}=\sqrt{C_0}(1+\theta)(t+t_0),~~~~~~t_0=const.
$$
For any deficit angle the core of dislocation is reached by the radial
extremal at finite time. It is naturally continued through the
conical singularity: we consider two radial extremals with angles
$\f$ and $\f+\pi$ as two halves of one complete extremal.
This is one possibility when the phonon is considered to go through the
conical singularity. The other possibility is to assume that the phonon
reaches the singularity in finite time and disappears. The present
model does not allow to choose between the alternatives, and one has
to invoke other theoretical or experimental arguments. This is beyond
the scope of the paper.

Note that circle extremals, $r=const$, are absent as the consequence
of equation (\ref{extrer}), though integrals
(\ref{extfin}), (\ref{extsin}) admit this solution. This happens
because to obtain the first integrals equations (\ref{extrer}),
(\ref{extref}) must be multiplied by $\dot{r}$.

The analysis of extremals shows that the motion of phonons is not a
potential one. That is, there is no effective potential yielding the
same trajectories in the flat Euclidean space. Indeed, the observed
kinetic energy (\ref{eobkie}) suggests the potential
$$
  U=-\frac12C_0r^{-2\theta}+const
$$
for the total energy to be conserved. Then one gets the Lagrangian
$$
  L=\frac12\dot r^2+\frac12r^2\dot\f^2+\frac12C_0r^{-2\theta}
$$
with Euclidean metric but nontrivial potential. This Lagrangian yields
qualitatively different trajectories for a particle.
\section{Cosmic strings and multiple images            \label{sangse}}
\begin{figure}[htb]
 \begin{center}
 \leavevmode
 \epsfysize=6cm
 \epsfbox{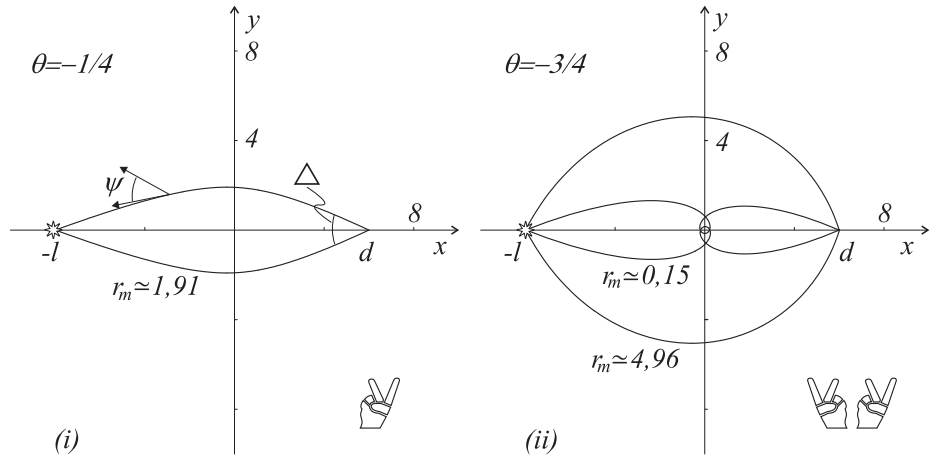}
 \end{center}
 \caption{Double images of a star located behind a cosmic string with
          small deficit angle {\it (i)}.
          If a deficit angle is close to $-2\pi$ then a star may have
          any even number of images. In Fig.{\it(ii)} we show a star
          with four images, two of them being produced by light rays
          making one turn around a cosmic string.
                                                      \label{fdouim}}
\end{figure}

The metric for a wedge dislocation with a negative deficit angle,
$\theta<0$,
coincides with the spatial part of the metric for a cosmic string
first found in \cite{Vilenk81}. It is interesting that the
metric for a cosmic string was obtained in a quite different way by
solving four-dimensional Einstein equations in the linear approximation.
Nevertheless the metric is essentially the same, and therefore
extremals found in the previous section describe also light rays
in the presence of a cosmic string. The well known effect of a
cosmic string is the appearance of double images of a star located
behind the cosmic string, see Fig.~\ref{fdouim}{\it (i)}.
In this section we derive a general formula for the angular separation
between the images. Originally, the angular separation was obtained
for small angles by calculating the angle in the coordinate system where
extremals are straight lines (that is, in the media with removed wedge but
not deformed). We obtain the formula which is valid for all angles
and in the coordinate system describing the media after the dislocation
is created.

Suppose that a star and an observer are located at distances $l$ and $d$,
respectively, as shown in Fig.~\ref{fdouim}.
The angle $\psi$ between the tangent vector to a light ray and the radial
direction is defined by the expression
$$
  \tan\psi=\frac{rd\f}{dr}.
$$
Using equation (\ref{eforwe}) one easily finds
$$
  \psi=-(1+\theta)(\phi+\phi_0),
$$
up to an addition of $2\pi n$, $|n|=1,2,\dots$ Location points of a
star and observer are defined by the equation
\begin{eqnarray}                                        \label{elocst}
  \left(\frac l{r_m}\right)^{2(1+\theta)}\sin^2\psi_1&=&1,
\\                                                      \label{elocob}
  \left(\frac d{r_m}\right)^{2(1+\theta)}\sin^2\psi_2&=&1,
\end{eqnarray}
where
\begin{eqnarray}                                        \label{epsion}
  \psi_1&=&-(1+\theta)(\pi+2m\pi+\f_0),
\\                                                      \label{epsitw}
  \psi_2&=&-(1+\theta)\f_0.
\end{eqnarray}
Here $m$ is the number of rotations of the light ray around the cosmic
string before reaching the observer. The whole number of rotations of
the corresponding extremal may be greater. The difference between
$\psi_2$ and $\psi_1$ must be less then $\pi$. Otherwise a star and an
observer would be crossed by different branches of an extremal.
Therefore the inequality
\begin{equation}                                        \label{emaxin}
  (1+\theta)\pi(1+2m)<\pi
\end{equation}
restricts the maximal number of rotations for a given deficit angle.
As the consequence of equations (\ref{epsion}) and (\ref{epsitw}) we have
$$
  \psi_2=\psi_1+\pi(1+\theta)(1+2m).
$$
Denoting the angular separation between images by
$\triangle=2(\pi-\psi_2)$
one easily deduces from (\ref{elocst}), (\ref{elocob}) a general formula
for the angular separation between two symmetric images produced by the
light rays making $m$ rotations around the string
\begin{equation}                                        \label{eangse}
  d^{(1+\theta)}\left|\sin\frac\triangle2\right|
  =l^{(1+\theta)}
  \left|\sin\left(\frac\triangle2+\pi(1+\theta)(1+2m)\right)\right|.
\end{equation}
Here $m$ is any natural number satisfying the inequality (\ref{emaxin}).
For small deficit angle, $|\theta|\ll1$, an extremal cannot make a
turn around the dislocation, $m=0$. Therefore there are only two
images separated by a small angle, $\triangle\ll1$. In this case
equation (\ref{eangse}) reduces to the well known formula \cite{VilShe94}
\begin{equation}                                        \label{eanses}
  \triangle=\frac l{d+l}|2\pi\theta|.
\end{equation}
The coincidence of the results is not trivial, because the angle is
computed in different coordinate systems. It means that the coordinate
transformation describing a defect creation must be described by a
conformal transformation which preserves angles. This is indeed the
case as will be shown in the next sections. In fact, this assumption was
made but not written in the original derivation of (\ref{eanses}).

Let us consider a general case (\ref{eangse}) in more detail. Light
rays emitted by a star are parametrized by the angle $\psi_1$.
They cross the $x$ axis at point $d$ defined by the equation
\begin{equation}                                        \label{edisob}
  \left(\frac dl\right)^{2(1+\theta)}
  \frac{\sin^2[\psi_1+(1+\theta)\pi(1+2m)]}{\sin^2\psi_1}=1.
\end{equation}
We see that for a given $m$ the observer at any distance $d$
sees two images: one above and one below the cosmic string as
shown in Fig.~\ref{fdouim}. If the deficit angle is sufficiently
close to $-2\pi$ then an extremal can make several turns around
the cosmic string. If the maximal number of rotations for which the
inequality (\ref{emaxin}) is satisfied equals $M$ then the observer
sees $2(1+M)$ images: two images caused by the light rays which do not
make a rotation before reaching the observer, two images because of
the extremals making one rotation, and so on up to $M$ rotations.
In Fig.~\ref{fdouim}{\it(ii)} we show a star with four images for
$\theta=-3/4$. For this deficit angle any extremal makes one full
rotation around the string but two of them reach the observer before
making a turn as shown in the picture. Light making an additional
rotation reaches an observer later and produces a time delay.
\section{Extremals for $N$ wedge dislocations          \label{sexwen}}
Let us consider an arbitrary number $N$ of parallel wedge dislocations,
characterized by deficit angles $\theta_n$, $n=1,\dots,N$. The corresponding
two-dimensional metric $dl^2$ in complex coordinates $z=x+i y$
has the form \cite{KatVol92}
\begin{equation}                                        \label{emettw}
  dl^2=dzd\bar z\prod_n[(z-z_n)(\bar z-\bar z_n)]^{\theta_n}.
\end{equation}
Here a bar denotes complex conjugation. Dislocations intersect
the $x,y$ plane in the points $z_n$. Direct integration of the
corresponding equations for extremals is sophisticated in this
coordinate system because in general there is no Killing vector
field in the $x,y$ plane. Even in the case of one edge dislocation
(the dipole of two wedge dislocations) the problem requires numerical
analysis if the metric is first expanded at large distances and then
extremals are found \cite{Moraes96}. However, the problem becomes
almost trivial if one does not expand the metric and note that there
is a coordinate system where the metric becomes Euclidean. In complex
coordinates denoted by $w$ it has the Euclidean form
\begin{equation}                                        \label{emeteu}
  dl^2=dwd\bar w,
\end{equation}
where
\begin{equation}                                        \label{ecapiz}
  w=u+iv=e^{i\al}\int^zd\z\prod_n(\z-z_n)^{\theta_n}+C.
\end{equation}
Here $\al$ and $C$ are arbitrary real and complex numbers.
The contour of integration
should not cross and contain closed loops around points $z_n$.
We discuss this integral in detail in Section \ref{sconma} in
several simple cases.
If all dislocations go through the $x$-axis, ${\rm Im}z_n=0$, then this
function is called the Christoffel--Schwarz integral (see,
for example, \cite{Evgraf91}) and is well known in complex analysis.
It conformally maps the upper half plane $z>0$ onto the polygon
with one vertex lying at infinity and other vertices at points
$w_n$. Each point $z_n$ is mapped into the vertex $w_n$ with
the inner angle $\pi(1+\theta_n)$. In a general case when
${\rm Im}z_n\ne0$ for some dislocations the conformal map becomes more
involved.

The integral (\ref{ecapiz}) yields the solution to the
Hamilton--Jacobi equation (\ref{ehamja})
$$
  S=\sqrt{2E}\sqrt{(w-c)(\bar w-\bar c)}-Et,~~~~~~c=a+i b=const.
$$
For the Euclidean metric (\ref{emeteu}) extremals are straight lines,
\begin{equation}                                        \label{eforme}
  (u-a)\sin\g=(v-b)\cos\g,
\end{equation}
or
\begin{equation}                                        \label{eformc}
  (w-c)e^{-i\g}=(\bar w-\bar c)e^{i\g}
\end{equation}
going through a point $(a,b)$ at an angle $\g$.
Scattering of a phonon in real crystal occurs in the original Cartesian
coordinates $x^\mu$. Therefore equation (\ref{eforme}) has to be written
in the original system. This cannot be done for arbitrary distribution
of defects because the integral (\ref{ecapiz}) cannot be taken in a
general case. To describe scattering of phonons one needs
only asymptotics at large distances $|z|\gg|z_n|$. In the following
sections we analyze the scattering up to the quadrupole expansion.
It is characterized by three global parameters of a distribution.

The total deficit angle (or the charge) normalized on $2\pi$ is
\begin{equation}                                        \label{etotan}
  \Theta=\sum_{n=1}^N \theta_n.
\end{equation}
We assume that $\Theta>-1$ for the physical reason because one 
cannot withdraw more than the whole media. The total Burgers vector
(the dipole momentum) and the quadrupole momentum normalized on
$2\pi$ are
\begin{eqnarray}                                        \label{etotbv}
  B&=&\sum_{n=1}^N \theta_n z_n,
\\                                                      \label{etotqm}
  M&=&\frac12\sum_{n,k=1}^N
  \left(\theta_n\theta_k-\theta_n\dl_{nk}\right) z_n z_k
  =\frac12B^2-\frac12\sum_n\theta_nz_n^2.
\end{eqnarray}
In this definition the Burgers vector and the quadrupole momentum should
be considered as complex numbers. In the next section we define the
Burgers vector as a vector. It has the same absolute value as
$B$ but differs from the vector joining the origin of the coordinate
system with the point $B$ in the complex plane.

The integral (\ref{ecapiz}) has different asymptotics depending
on the total deficit angle. We write down the first three terms for
different values of $\Theta$.
\begin{eqnarray}                                        \label{easfir}
  w&\approx&e^{i\al}\left(
  \frac1{\Theta+1}z^{\Theta+1}-\frac B\Theta z^\Theta
  +\frac M{\Theta-1}z^{\Theta-1}\right)+C,~~~~
  \Theta\ne0,1,
\\                                                      \label{eassec}
  w&\approx&e^{i\al}\left(z-B\ln z-\frac Mz\right)+C,
  ~~~~~~~~~~~~~~~~~~~~~~~~~~\Theta=0,
\\                                                      \label{easthi}
  w&\approx&e^{i\al}\left(\frac12z^2-Bz+M\ln z\right)+C,
  ~~~~~~~~~~~~~~~~~~~~~~\Theta=1.
\end{eqnarray}
In the case $\Theta=0$ no wedge of media is added or removed.
The values $\Theta=1,2,\dots$ determine the position of $\ln$-term
in the expansion of $w(z)$. The quadrupole term is logarithmic for
$\Theta=1$.

Expansions (\ref{easfir})--(\ref{easthi}) show that any extremal going
to infinity in the $z$-plane goes also to infinity in the $w$-plane.
In the latter case extremals are obviously complete, and thus they are
complete in the $z$-plane. If an extremal does not go to infinity
then it is also complete because it is either closed or it can be
continued to infinite value of the canonical parameter within a finite
domain. Note that we have only conical singularities and extremals
can be naturally continued through them.

If the total angle $\Theta\ne0$ then the behavior at large distances is
characterized by the first term in equation (\ref{easfir}) describing
single wedge dislocation discussed in the previous section. Therefore
we consider the dipole and quadrupole approximation in the following
sections.
\section{Conformal maps and dislocations               \label{sconma}}
The integral (\ref{ecapiz}) provides a conformal map between
a plane with arbitrary number of conical singularities corresponding to
wedge dislocations ($z$-plane) and a Riemannian surface (covered by
$w$-coordinates) with the Euclidean metric. This map has a clear physical
interpretation corresponding to a defect creation. One starts with the
Euclidean plane, then cuts or adds parts of the plane making a
Riemannian surface with boundaries. Then the boundaries are glued together.
The gluing process is uniquely defined by the conformal map. The
points on a boundary are considered as the same if they correspond to
the same point on the cut in the $z$-plane. Below we consider several
examples of conformal maps defining the defects considered in the
present paper.

The Christoffel--Schwarz integral depends on two
constants. We suppose that the point $z_1$ always
coincides with the origin of the coordinate system in the $z$-plain,
and no point $z_n$, $n=2,3,\dots,$ lies on the negative half of the
real axis. Then the constants may be chosen in such a way that the
point $w_1$ coincides with the origin in the $w$-plain and negative
part of $x$-axis is mapped onto negative part of the $u$-axis.

\begin{figure}[htb]
 \begin{center}
 \leavevmode
 \epsfxsize=12cm
 \epsfbox{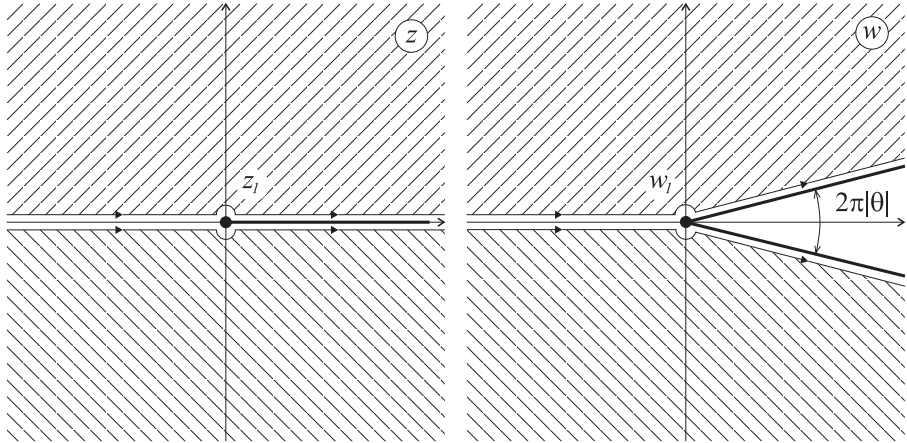}
 \end{center}
 \caption{The Christoffel--Schwarz conformal map for a wedge dislocation.
                                                       \label{fwedgec}}
\end{figure}

The simplest example of the Christoffel--Schwarz map (\ref{ecapiz})
is provided by the wedge dislocation discussed in Section \ref{swedge}.
In this case we have only one dislocation core located in the
origin of the coordinate system
$$
  z_1=0,~~~~~~\theta_1=\theta.
$$
The corresponding conformal map is given by the function
\begin{equation}                                        \label{ewecmf}
  w=\frac{e^{-i\pi\theta}}{1+\theta}z^{1+\theta},
\end{equation}
and is shown in Fig.~\ref{fwedgec}.
The Christoffel--Schwarz integral maps the upper half plane into the
polygon with two vertices. One vertex is located at infinity while the
other coincides with the origin $w_1=0$ and has inner angle $\pi(1+\theta)$.
The whole $z$-plane with a cut along the positive half of the real axis is
mapped onto the $w$-plane with the wedge of deficit angle $2\pi\theta$
removed or added to the plane for $-1<\theta<0$ or $0<\theta$, respectively.
This conformal map reflects the process of the creation of a wedge
dislocation. One takes an infinite elastic media without a defect,
cuts out the wedge $2\pi|\theta|$ for $-1<\theta<0$, and identifies the
points corresponding to the same point on the cut in the $z$-plane.
For $0<\theta$ one has to cut the $w$-plane, move apart the sides, and
insert the wedge of a media without stresses.

\begin{figure}[htb]
 \begin{center}
 \leavevmode
 \epsfxsize=12cm
 \epsfbox{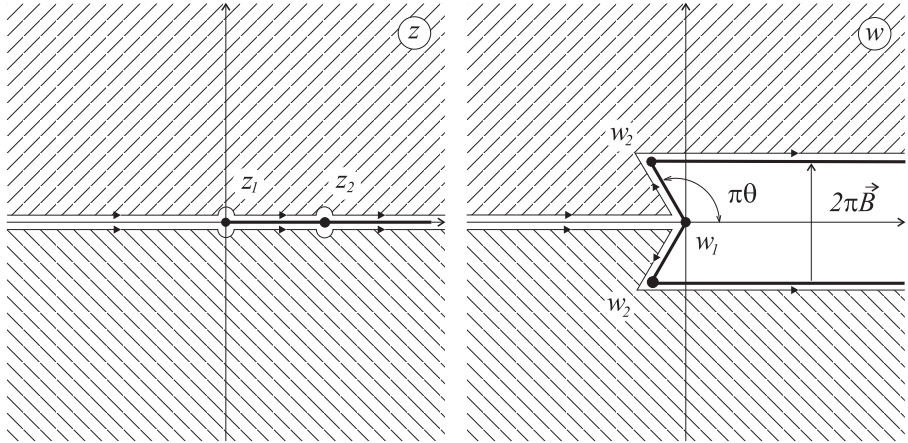}
 \end{center}
 \caption{The Christoffel--Schwarz conformal map for a
          dipole of two wedge dislocations representing one
          edge dislocation.
                                                       \label{fedgcm}}
\end{figure}

The dipole of two wedge dislocations is called an edge dislocation and
is characterized by a constant Burgers vector $\vec B$ to be defined later.
To construct the conformal map we choose the following locations and
angles of the wedge dislocations
\begin{equation}                                        \label{edipdi}
\begin{array}{lll}
  z_1=0, & & \theta_1=-\theta,
\\
  z_2=h, & & \theta_2=\theta.
\end{array}
\end{equation}
We assume that $0<\theta<1$.
Then the Christoffel--Schwarz integral (\ref{ecapiz}) takes the form
\begin{equation}                                        \label{ecsind}
  w=\int_0^zd\z\left(\frac{\z-h}\z\right)^\theta.
\end{equation}
Here the constants are chosen in such a way that $w_1(z_1)=0$.
Location of the second vertex in the upper half plain
(see Fig.~\ref{fedgcm})
is expressed in terms of the gamma-functions
\begin{equation}                                        \label{esecwe}
  w_2=\int_0^hd\z\left(\frac{\z-h}\z\right)^\theta
  =he^{i\pi\theta}\int_0^1dx\left(\frac{1-x}x\right)^\theta
  =he^{i\pi\theta}\G(1-\theta)\G(1+\theta).
\end{equation}
In Fig.~\ref{fedgcm} we show the conformal map for $1/2<\theta<1$. The
$z$-plane has a cut along the positive part of the real axis. The upper
half plane is mapped onto the triangle in the upper half of the $w$-plain
with one vertex lying at infinity. The lower half of the $z$-plain is
mapped symmetrically. The whole $z$-plain with the cut is mapped onto
the $w$-plain with the removed strip as shown in the picture.
We define the Burgers vector as the vector joining two points
$w_-$ and $w_+$ corresponding to lower and upper shores of the cut
in the $z$-plane. For an edge dislocation for $x>z_2$ it is a constant
vector. The absolute value of the normalized vector is
\begin{equation}                                        \label{edefbv}
  2\pi|\vec B|=w_+-w_-=2{\rm Im}w_2.
\end{equation}
This definition is applicable for arbitrary distribution of wedge
dislocations. In a general case the cut may be defined as the line
joining the sequence of points $z_1,\dots,z_N,\infty$ including the
infinite point. Of course, the corresponding Burgers vector may not
be constant. Using the expression (\ref{esecwe}) and the property of
gamma-functions we get for an edge dislocation
$$
  2\pi|\vec B|=2h\G(1-\theta)\G(1+\theta)\sin(\pi\theta)=2h\pi\theta.
$$
This remarkable result shows that the normalized Burgers vector equals
the product $h\theta$. Note that this is an exact result for a
dipole of two wedge dislocations, and it coincides with the definition
(\ref{etotbv}) for the distribution (\ref{edipdi}) used in the large
distance expansion of the Christoffel--Schwarz integral.

\begin{figure}[htb]
 \begin{center}
 \leavevmode
 \epsfxsize=12cm
 \epsfbox{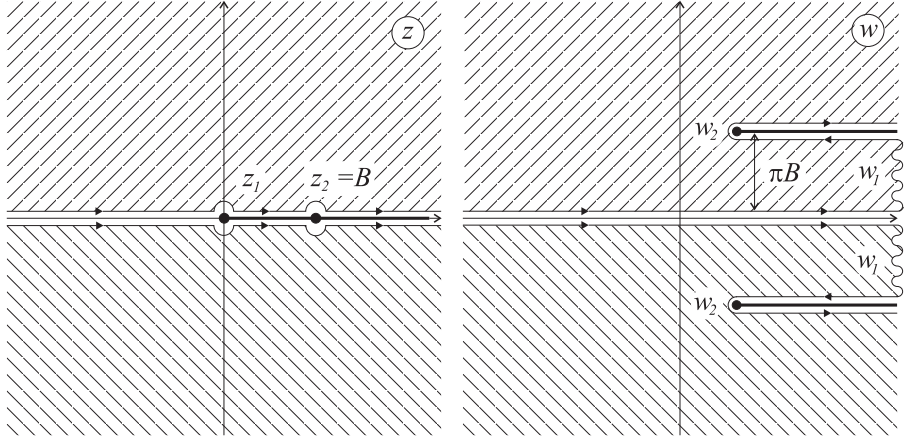}
 \end{center}
 \caption{The conformal map in the dipole approximation (an edge
          dislocation).
                                                       \label{fdipcm}}
\end{figure}

Let us compare the conformal map for a dipole of wedge dislocations with
the conformal map (\ref{eassec})
of the pure edge dislocation $\Theta=0$, and $B\ne0$
\begin{equation}                                        \label{edipap}
  w=z-B\ln z+i\pi B,
\end{equation}
where we have specified the values of the constants. Elementary analysis
yields the conformal map shown in Fig.~\ref{fdipcm}.
It maps the upper half plane of the $z$-plain onto the upper half plain
of the $w$-plain with the cut as shown in the picture. The origin of
the coordinate system is mapped into the infinite point, $w_1(z_1)=\infty$.
This conformal map has less clear physical interpretation but is more
universal because it does not depend on the details of the distribution
at small distances.

\begin{figure}[htb]
 \begin{center}
 \leavevmode
 \epsfxsize=12cm
 \epsfbox{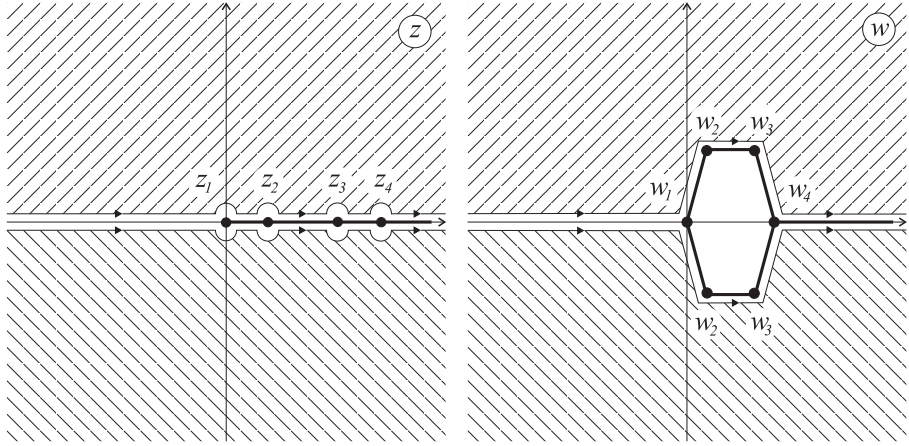}
 \end{center}
 \caption{The conformal map for the four wedge dislocations
          representing the quadrupole source.
                                                       \label{fourcm}}
\end{figure}

The pure quadrupole distribution of wedge dislocations may be organized
in the following way
\begin{eqnarray*}
  z_1=0, && \theta_1=-\theta,
\\
  z_2=h, && \theta_2=\theta,
\\
  z_3=l, && \theta_3=\theta,
\\
  z_4=l+h, && \theta_4=-\theta,
\end{eqnarray*}
where $h,l$, and $\theta$ are constants. For definiteness we assume
that $l>h$ and $0<\theta<1$. Then only the quadrupole momentum differs
from zero
$$
  \Theta=0,~~~~~~B=0,~~~~~~M=\theta lh.
$$
The corresponding conformal map is provided by the
Christoffel--Schwarz integral
\begin{equation}                                        \label{eChScq}
  w=\int_0^zd\z\left[\frac{(\z-h)(\z-l)}{\z(\z-l-h)}\right]^\theta.
\end{equation}
It cannot be taken in elementary functions but may be analyzed
qualitatively. The location of the vertices in the upper half of the
$w$-plain is given by the convergent integrals
\begin{eqnarray*}
  w_2&=&he^{i\pi\theta}\int_0^1dx
  \left[\frac{(1-x)(p-x)}{x(p+1-x)}\right]^\theta,
\\
  w_3&=&w_2+(l-h)\int_0^1dx
  \left[\frac{x(1-x)}{(q+x)(q+1-x)}\right]^\theta,
\\
  w_4&=&w_3+he^{-i\pi\theta}\int_0^1dx
  \left[\frac{(1-x)(p-x)}{x(p+1-x)}\right]^\theta,
\end{eqnarray*}
where
$$
  p=\frac lh>1,~~~~~~q=\frac h{l-h}>1.
$$
We see that ${\rm Im}w_4=0$. The lower half plain is mapped
symmetrically. The corresponding conformal map is shown
in Fig.~\ref{fourcm}.
This conformal map has evident physical interpretation. The
dislocation is made by withdrawing a hexagonal tube of media.

\begin{figure}[htb]
 \begin{center}
 \leavevmode
 \epsfxsize=12cm
 \epsfbox{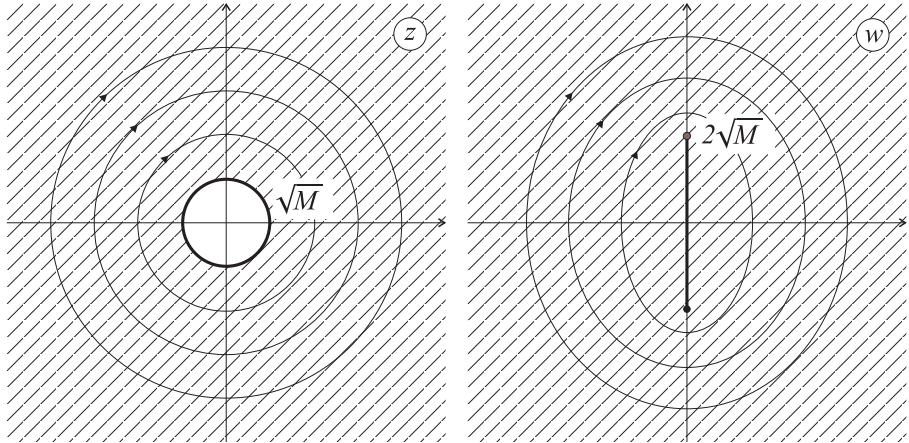}
 \end{center}
 \caption{The conformal map in the quadrupole approximation.
                                                       \label{fquacm}}
\end{figure}

In the pure quadrupole approximation, $\Theta=0$, $B=0$, the conformal
map for arbitrary distribution of defects (\ref{eassec}) takes the
form
\begin{equation}                                        \label{equacm}
  w=z-\frac Mz,
\end{equation}
where the constants have been specified. It maps circles into ellipses as
shown in Fig.~\ref{fquacm}.
The circles of radius greater than $\sqrt M$
are mapped preserving orientation, and the outer part of the $z$-plain,
$|z|>\sqrt M$ is mapped onto the whole $w$-plain with the cut along the
imaginary axis. The inner circles are mapped with opposite orientation,
and the interior $|z|<0$ also covers the whole $w$-plain. We see that
the function (\ref{equacm}) maps the whole $z$-plain onto a Riemann
surface consisting of two copies of $w$-plain glued together along
the cut. We leave it to the reader to imagine how this conformal map
reflects the defect creation.
\section{The dipole approximation                      \label{sdipap}}
\begin{figure}[htb]
 \begin{center}
 \leavevmode
 \epsfxsize=12cm
 \epsfbox{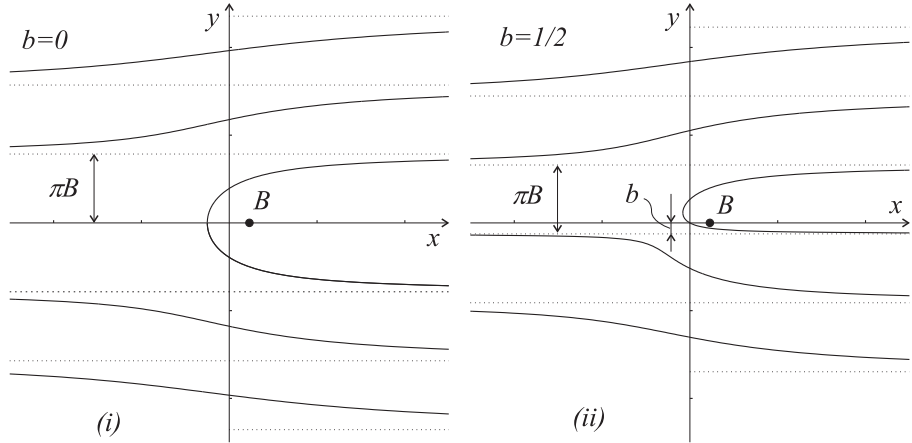}
 \end{center}
 \caption{Extremals for the edge dislocation for $b=0$ {\it (i)} and
          $b=1/2$ {\it (ii)} which are perpendicular to the Burgers
          vector at infinity. For any $b\ne0$ the returning extremal
          goes through the core of dislocation.
          \label{fedexh}}
\end{figure}

If the total deficit angle equals zero, $\Theta=0$, and $B\ne0$ then
the scattering at large distances is described by the dipole term
in (\ref{eassec}). The line element corresponding to the conformal
transformation (\ref{edipap}) in polar coordinates is asymptotically
flat
\begin{equation}                                        \label{eliedi}
  dl^2=\left(1-B\frac{2\cos\f}r\right)(dr^2+r^2d\f^2).
\end{equation}
The extremals (\ref{eforme}) have the form
\begin{equation}                                        \label{extdip}
  r\sin(\f-\g)+B(\ln r\sin\g+\f\cos\g)-a\sin\g+b\cos\g=0.
\end{equation}
In the dipole approximation the scattering is characterized by the
vector, and trajectories depend on the angle at which extremals go to
infinity. We consider two cases. For $\g=0$ the trajectories at
infinities are perpendicular to the Burgers vector. In this case
equation (\ref{extdip}) reduces to
$$
  \frac yx=-\tan\frac{y+b}B,
$$
and extremals are parametrized by one arbitrary constant $b$.
In figure~\ref{fedexh} extremals are shown for $b=0$ and $b>0$,
respectively. Extremals for $b<0$ are obtained from those depicted
in Fig.~\ref{fedexh} {(\it ii)} by the reflection $y\rightarrow-y$.
Extremals have asymptotes at infinity shifted by $\pi B$. So the net
result of the scattering is the shifting of a phonon trajectory defined
by the Burgers vector and a time delay. Note the existence of
returning trajectories near the dislocation when phonons move from the
right. If phonons move from the left there is no backward scattering.
It means that an edge dislocation is "invisible" when it is being seen
perpendicular to the Burgers vector from the left.

\begin{figure}[htb]
 \begin{center}
 \leavevmode
 \epsfxsize=12cm
 \epsfbox{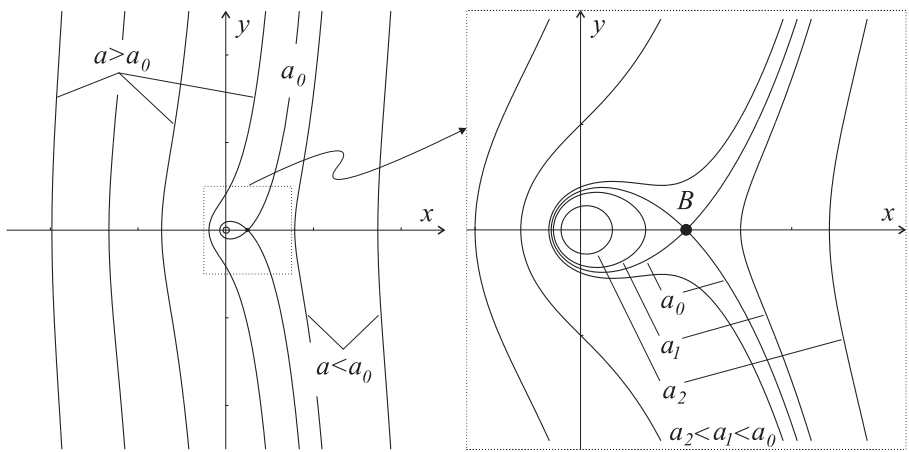}
 \end{center}
 \caption{Extremals for an edge dislocation which are
          parallel to the Burgers vector at infinity.
          They have no asymptotes. For $a<a_0$ an extremal has two
          branches.                                    \label{fedexv}}
\end{figure}

The extremals which are parallel to the Burgers vector at infinity
have different behavior and are shown in Fig.~\ref{fedexv}.
They correspond to $\g=\pi/2$ and are defined by the equation
\begin{equation}                                        \label{ehoexd}
  x^2+y^2=\exp\left(2\frac{x+a}B\right).
\end{equation}
In this case they are parametrized by one arbitrary constant $a$.
At infinity the extremals have no asymptotes. This means that even at
infinity the observer can feel the dislocation because trajectories are
not straight lines. There is also a time delay.
For $a=a_0$ where
$$
  a_0=B\ln B
$$
the extremal has the intersection point at $(B,0)$. For $a>a_0$
an extremal has one branch. If $a<a_0$ then solution of equation
(\ref{ehoexd}) has two branches. One branch starts and ends at infinity
while the other is closed and surrounds the dislocation core.
In fact this branch may not be closed if the velocity along the $x^3$
axis is nonzero. In that case it looks like a spiral.

Let us remind the reader that extremals are drawn for the dipole
approximation, and therefore one may expect qualitative agreement
at large distances from the core of dislocation.

Extremals for an edge dislocation were analyzed numerically in
\cite{Moraes96}. Qualitative behavior is the same except for the loss of
returning trajectories in Fig.~\ref{fedexh} and closed trajectories in
Fig.~\ref{fedexv}.
\section{The quadrupole approximation                  \label{squdap}}
\begin{figure}[htb]
 \begin{center}
 \leavevmode
 \epsfxsize=12cm
 \epsfbox{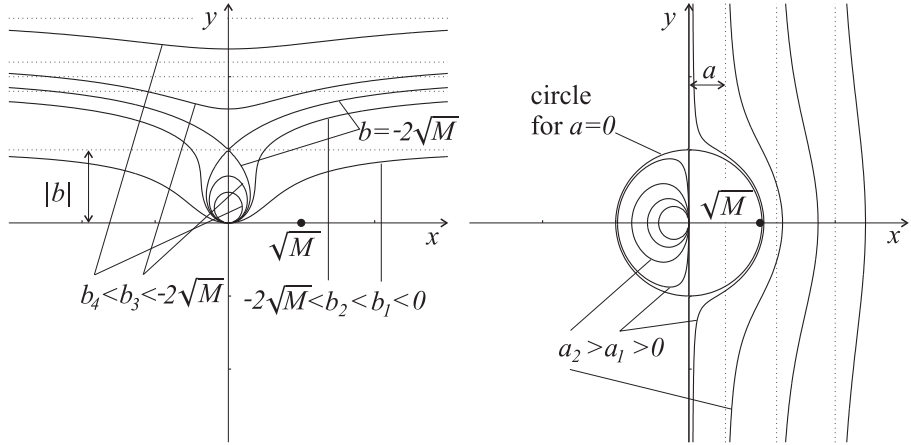}
 \end{center}
 \caption{Extremals in the quadrupole approximation which are
          parallel to $x-$ {\it (i)} and $y$-axis {\it (ii)} at
          infinity. An extremal has two branches for
          $b<-2\protect\sqrt M$ and arbitrary $a$.     \label{fquade}}
\end{figure}

In the quadrupole approximation, $\Theta=0,~B=0$, $M\ne0$, the
conformal map (\ref{eassec}) at large distances reduces to
\begin{equation}                                        \label{equapp}
  w=z-\frac Mz.
\end{equation}
Without loss of generality we set ${\rm Im} M=0$ and suppose that
$M>0$. To get the scattering for negative $M$ one has to turn the
whole picture on the angle $\pi$. The line element corresponding to
the conformal map (\ref{equapp}) has the form
\begin{equation}                                        \label{eliequ}
  dl^2=\left(1+2M\frac{\cos(2\f)}{r^2}\right)(dr^2+r^2d\f^2)
\end{equation}
and goes to the Euclidean line element faster than in the dipole
approximation as $r\rightarrow\infty$. In the quadrupole approximation
equation (\ref{eforme}) for an extremal reduces to
\begin{equation}                                        \label{extqud}
  r\sin(\f-\g)+\frac Mr\sin(\f+\g)-a\sin\g+b\cos\g=0.
\end{equation}
At infinity $r\rightarrow\infty$ this equation coincides with that for
a straight line, and therefore any extremal has asymptote.
Let us consider two cases. Extremals parallel to the
$x$ axis at infinity correspond to $\g=0$ and are defined by the
equation
\begin{equation}                                        \label{equadx}
  x^2+y^2=-\frac{My}{y+b}.
\end{equation}
There is no extremal for $b=0$. For negative $b<0$ all extremals lie
in the upper half plane and are shown in Fig.~\ref{fquade}{\it(i)}.
Extremals for $b>0$ are obtained from those for $b<0$ by reflection
$y\rightarrow-y$. For $-2\sqrt M<b<0$ an extremal has only one branch
going from $x=-\infty$ to $x=+\infty$ and touching the
$x$-axis at the origin of the coordinate system. If $b=-2\sqrt M$, then
the extremal has the intersection point at $x=0$, $y=\sqrt M$. For
$b<-2\sqrt M$ an extremal consists of two branches. One infinite
branch goes from $x=-\infty$ to $x=+\infty$ and does not touch the
$x$-axis. The other branch is closed and goes through the origin.
Any infinite branch has the same asymptote $y=-b$ at both sides as
$x\rightarrow\pm\infty$, and the scattering reduces to a time delay.

The extremals parallel to the $y$ axis at infinity correspond to
$\g=\pi/2$ and are defined by the equation
$$
  x^2+y^2=\frac{Mx}{x-a}.
$$
It differs from (\ref{equadx}) by the sign in the right hand side and
the exchange $x\leftrightarrow y$. Here $M$ is also supposed to be
positive.  For $a=0$ the extremal is the circle of radius $\sqrt M$.
Extremals with positive $a>0$ are shown in Fig.~\ref{fquade}{\it(ii)}.
Extremals corresponding to negative $a<0$ are obtained by the
reflection $x\rightarrow-x$. Any extremal consists of two branches.
One branch is infinite and goes from $y=-\infty$ to $y=+\infty$ with the
same asymptote $x=a$. The other branch is closed and located within the
circle on the other half of the plane. It goes through the origin.
So the scattering reduces only to a time delay.
\section{Continuous distribution of defects            \label{scondi}}
One of the main advantages of geometrical description of defects is the
possibility to describe continuous distributions of dislocations and
disclinations. First, we consider circularly symmetric distribution of
wedge dislocations. They are assumed to be uniformly distributed on the
disk of radius $R$ with the following density of deficit angles
$$
  \rho(r)=\left\{\begin{array}{ll}q, & r\le R,\\ 0, &r>R. \end{array}\right.
$$
The normalized total deficit angle for this distribution is
$$
  \Theta=\frac12q R^2.
$$
The conformal factor of the corresponding metric
\begin{equation}                                        \label{econfo}
  g_{\al\bt}=e^{2\phi}\dl_{\al\bt}
\end{equation}
satisfies the three-dimensional Einstein equations which in the
considered case reduce to the Poisson equation \cite{KatVol92}
$$
  \triangle\phi=-\rho(r)
$$
and is defined by the integral
$$
  \phi(r)=\frac1{2\pi}\int d\vec s\rho(\vec s)\ln|\vec s-\vec r|.
$$
The result is
\begin{equation}                                        \label{elogpo}
  \phi(r)=\left\{\begin{array}{ll}
  {\displaystyle\frac q2} R^2\ln R-
  {\displaystyle\frac{qR^2}4}+
  {\displaystyle\frac{qr^2}4}, & r\le R, \\ \\
  {\displaystyle\frac q2} R^2\ln r, & r>R.
                                \end{array}\right.
\end{equation}
Outside the dislocations the line element has the form
$$
  dl^2=r^{qR^2}(dr^2+r^2d\varphi^2).
$$
This means that the distribution of wedge dislocations as seen from the
outside is the same as for one wedge dislocation (\ref{emetwe}) with
the deficit angle $\Theta$. Therefore trajectories of phonons at large
distances are the same as for a single wedge dislocation. This is the
result one would not expect because Einstein equations are nonlinear,
and, in general, there is no superposition principle.
Note that metric (\ref{econfo}) with the conformal factor (\ref{elogpo})
is the exact solution to the Einstein equations.

Now we consider the continuous distribution of edge dislocations
characterized by the density of the normalized Burgers vector
$\vec\bt(\vec r)$$=(\bt_x,\bt_y)$.
It is assumed to be directed along the $x$-axis, $\bt_y=0$, and uniformly
distributed on the disk of radius $R$
$$
  \bt_x=\left\{\begin{array}{ll} \bt, & r\le R, \\
                              0,   & r>R.   \end{array}\right.
$$
In this case the Einstein equations also reduce to the Poisson
equation for the conformal factor
$$
  \triangle\phi=2\pi(\vec\nabla\vec\bt).
$$
Its solution has the form
\begin{equation}                                        \label{ecodib}
  \phi=\left\{\begin{array}{ll}
  {\displaystyle\frac{\bt r\cos\f}2},      & r<R, \\ \\
  {\displaystyle\frac{\bt R^2\cos\f}{2r}}, & r>R.\end{array}\right.
\end{equation}
At large distances, $r\gg R$, the space is described by the line element
(\ref{eliedi}) with
\begin{equation}                                        \label{ebucon}
  B=\bt\pi R^2,
\end{equation}
and is asymptotically flat.
We see that at large distances the continuous distribution of edge
dislocations behaves like one edge dislocation with the total
Burgers vector (\ref{ebucon}). The superposition principle is valid
only for the main term in the line element. There are further corrections
which distinguish one edge dislocation from the continuously distributed
ones. The first correction may be easily obtained in complex coordinates.
The conformal factor (\ref{ecodib}) outside the distribution yields the
line element
$$
  dl^2=\exp\left(-B\left(\frac1z+\frac1{\bar z}\right)
  \right)dzd\bar z.
$$
It may be written in flat form $dl^2=dwd\bar w$, where
$$
  w=\int^z d\z\exp\left(-\frac B\z\right)
  \simeq z-B\ln z-\frac{B^2}{2z}.
$$
This expansion yields the quadrupole correction, $M=B^2/2$, to the
continuous distribution of edge dislocations.
\section {Conclusion}
In the geometrical theory of defects an elastic media
is assumed to be a manifold equipped with a Riemann--Cartan
geometry, curvature and torsion tensors describing the surface densities
of disclinations and dislocations, respectively. When disclinations are
absent then curvature is equal to zero and the corresponding
$SO(3)$-connection can be locally gauged away. Then geometry is
entirely defined by the triad satisfying the three-dimensional
Einstein equations. For a given triad (or a metric) one has two choices:
either to consider a manifold as a Riemannian one with Christoffel's
symbols as a connection and nontrivial curvature or to introduce
a torsion in such a way that curvature corresponding to the metrical
connection is identically zero (a manifold with absolute parallelism
if one excludes paths around the singularities).
In the last case the manifold has nontrivial torsion defined by the
triad only. In this sense three-dimensional gravity and geometric
theory of dislocations differ but the solution to one problem yields
immediately the solution to the other and vice versa. That is a
solution to the three-dimensional Einstein equations may be always
interpreted as some distribution of defects in elastic media.
The correspondence between the notions is as follows:
\begin{center}
\begin{tabular}{rcl}
  spinless point particle&=&wedge dislocation \\
  point particle with spin&=&screw dislocation \\
  static point particles&=&parallel dislocations \\
  moving point particles&=&non parallel dislocations.
\end{tabular}                                                 \newline
\end{center}
Recently solutions for moving particles with and without spins were found
in three-dimensional gravity [35-38].
\nocite{BeCiVa95A,BeCiVa95B,BeCiVa96,CiaVal97}
It means that the Euclidean version of this solution describes an arbitrary
number of arbitrary oriented wedge and screw dislocations. This problem
is so complicated in the framework of ordinary elasticity theory that it
can hardly be solved. At the same time it can be solved in the framework
of the geometric theory of defects.

Another advantage of the theory of defects concerns the propagation of
elastic waves in the presence of defects considered in the paper. We are
not aware how to formulate this problem unambiguously even for a single
dislocation in the framework of elasticity theory. At the same time the
geometrical approach adopted in the present paper provides an almost standard
and simple answer: the perturbations propagate along extremals for a metric
describing given distribution of dislocations. In this way we have found
trajectories of phonons. Their behavior is quite different from that in
the potential motion of point particles. We hope that this non-potential
motion of phonons in the presence of defects may be confirmed or
refuted experimentally in the future. Maybe elastic media provides a
much better experimental field for testing gravity models.

In the present paper we have shown that a star located behind
a cosmic string may produce any even number of images. The number
of images is larger when the deficit angle is closer to $-2\pi$.
Analogous behavior of extremals is encountered also in general
relativity already for the Schwarzschild solution \cite{Chandr83}.
This effect leads to the appearance of multiple images of a star
located behind the singularity and changes the usual description
of gravitational lensing. If the singularity is strong enough
then an observer sees not two but a large even number of images of
a star. There is a wide field of speculations about what we really 
see in the sky.

This work is supported by the Russian Foundation for Basic Research,
Grants RFBR-96-010-0312 and RFBR-96-15-96131.

\end{document}